\documentclass[twocolumn,preprintnumbers,amssymb,floatfix,aps,prl]{revtex4}
\usepackage{graphicx}
\usepackage{dcolumn}
\usepackage{bm}

\begin{document}

\title{Microscopic Model for Photoinduced Magnetism in the Molecular Complex 
$[Mo(IV)(CN)_2(CN-CuL)_6]^{8+}$ Perchlorate}

\author{Rajamani Raghunathan$^\S$, S. Ramasesha$^\S$, 
Corine Mathoni\`ere$^\dagger$, Val\'erie Marvaud$^\ddagger$
\footnote{E-mail: rajamani@sscu.iisc.ernet.in; 
ramasesh@sscu.iisc.ernet.in; mathon@icmcb-bordeaux.cnrs.fr; 
marvaud@ccr.jussieu.fr}}

\affiliation {$^\S$Solid State and Structural Chemistry Unit, 
Indian Institute of Science, Bangalore 560 012, India\\
$^\dagger$Institut de Chimie de la Mati\`ere Condens\'ee de 
Bordeaux CNRS UPR 9048
87, Avenue du Docteur Schweitzer 33608 Pessac cedex (France)\\
$^\ddagger$Labaratoire de Chimie inorganique et Mat\'eriaux Mol\'eculaires
CNRS, UMR 7071, Universit\'e Pierre et Marie Curie, 75252 Paris Cedex 05 
(France)}


\begin{abstract}
A theoretical model for understanding photomagnetism in the heptanuclear
complex $[Mo(IV)(CN)_2(CN-CuL)_6]^{8+}$ perchlorate is developed. It is a
many-body model involving the active orbitals on the transition metal ions.
The model is exactly solved using a valence bond approach. The ground state
solution of the model is highly degenerate and is spanned by five S=0 states,
nine S=1 states, five S=2 states and one S=3 state. The orbital occupancies in
all these states correspond to six $Cu(II)$ ions and one diamagnetic $Mo(IV)$ ion. 
The optically excited charge-transfer (CT) state in each spin sector occur at 
nearly the same excitation energy of 2.993 eV for the physically
reasonable parameter values. The degeneracy of the CT states is largest in
the S=3 sector and so is the transition dipole moment from the ground state
to these excited states. Thus laser irradiation with light of this energy
results in most intense absorption in the S=3 sector. The life-time of the 
S=3 excited states is also expected to be the largest as the number of 
states below that energy is very sparse in this spin sector when compared to 
other spin sectors.  These twin features of our model explain the observed 
photomagnetism in the $[Mo(IV)(CN)_2(CN-CuL)_6]^{8+}$ complex.

\centerline{(Received date : \today)}
\end{abstract}

\maketitle

\section{I. Introduction}
In the expanding field of molecular magnetism, photomagnetism has 
considerable import as it affords the possibility of magnetic 
switching with the aid of light. In the photomagnetic systems that
we have studied here, the system before shining light is essentially 
paramagnetic, 
with negligible exchange interaction between the magnetic ions. 
Shining a burst of light of appropriate wavelength takes the 
system to a state in which there is a superexchange 
interaction between the paramagnetic ions. 
The photo-induced high-spin state is long lived at low 
temperatures, however, warming the system takes it back to 
the paramagnetic ground state \cite{takagi}. There also exist extended
systems in which the photoinduced exchange mechanism gives
rise to magnetic ordering; this unusual photo-effect is called 
the Photo Induced Phase Transition (PIPT) \cite{extendedCuMo}.  

The above phenomenon is quite distinct from the spin state 
transition observed in many inorganic complexes, wherein 
shining light leads to change in the ratio of the population 
of the low-spin (LS) state to that in the high-spin (HS) state. This 
phenomenon called the Light Induced Excited Spin State Trapping 
(LIESST), has been extensively studied in spin crossover 
complexes \cite{sato-acr} . These systems lie very close to the spin crossover 
point in the Tanabe-Sugano diagram \cite{tanabe-sugano}. When the system is excited 
by light, the non-radiative decay from the excited state occurs 
through intersystem crossings to the two nearly degenerate low 
energy states of different spin \cite{gutlich}. To recover the pure low-spin 
ground state, it would become necessary to warm the system and 
cool it to low temperatures. This phenomenon does not involve
any magnetic exchange interactions between the transition metal
ions.

Photomagnetism was experimentally first demonstrated in Fe-Co prussian 
blue complex K$_{0.2}$Co$_{1.4}$Fe(CN)$_6 \cdot$6.9H$_2$O.
This system is ferrimagnetic with a T$_c$ of 16K \cite{sato-science}.
Upon irradiation, the T$_c$ increased
to 19K. Besides, at fixed magnetic field and fixed temperature, the 
net magnetization increased upon irradiation. This increase in T$_c$ and 
magnetization are attributed to electron transfer from the diamagnetic 
$Fe(II)$ to the diamagnetic $Co(III)$ brought about by irradiation resulting 
in $Fe(III)$ ion with spin 1/2 and $Co(II)$ ion with spin 3/2 \cite{sato-PRB}. 
These paramagnetic ions,
besides contributing to an increase magnetization, also establish pathways 
for exchange which were blocked in the diamagnetic state, thereby increasing 
the T$_c$.

\begin{figure}
\includegraphics[height=3.2in,width=3.5in]{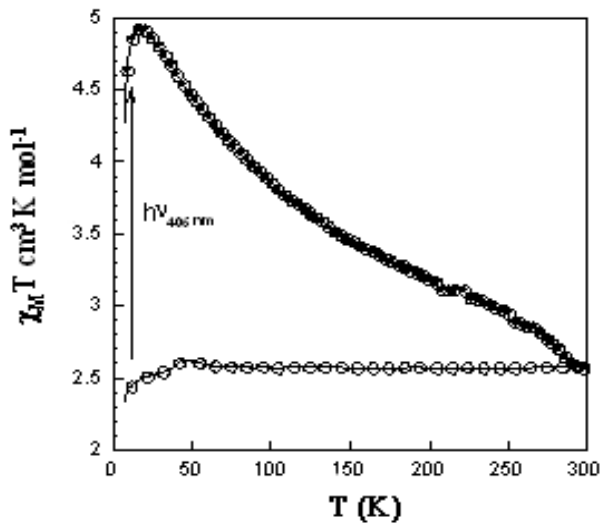}
\caption[]{\label{chit-expt} Temperature dependence of $\chi_MT$, 
($\circ$) before irradiation, ($\bullet$)
after irradiation.}
\end{figure}

The system we have studied theoretically is the $Mo-Cu$ molecular system. 
While photoinduced ferromagnetism has been observed in extended $Mo-Cu$ 
systems \cite{extendedCuMo}, we focus our attention in this paper on the 
molecular system $[Mo(IV)(CN)_2(CN-Cu(II)L)_6]^{8+}$ with perchlorate as 
the counter ions \cite{corine}. In the ground state, each molecule consists 
of six noninteracting spins, derived from the six $Cu(II)$ ions; the 
$Mo(IV)$ ion is diamagnetic. The magnetic behaviour of this molecule in the 
ground state corresponds to that of a system with six independent spin-1/2 
objects per formula unit with $\chi_MT$ value of 2.48 cm$^3$ K mol$^{-1}$ 
at 20K. On irradiation with blue light (406-415 nm) at 10K, the $\chi_MT$ 
value increases to 4.8 cm$^3$ K mol$^{-1}$. On turning off the light and 
warming the system, the $\chi_MT$ value peaks at 5 cm$^3$ K mol$^{-1}$ and 
decreases thereafter gradually to a room temperature value of 2.6 cm$^3$ K 
mol$^{-1}$. 
The experimental data after irradiation is in agreement with 75\% 
of the molecules in S=3 state with the remaining in the state corresponding 
to six isolated S=1/2 spins per molecule. Fig. \ref{chit-expt} 
shows the $\chi_MT$ vs $T$ plot of the system before and after irradiation.

In this paper, we develop a microscopic quantum many-body model to 
understand the photomagnetic behaviour of this system. In the next 
section, we describe our model and the computational scheme in detail. 
In the third section we present the results from our model. We show 
that the eigenvalue spectrum of the model can be classified according 
to the occupancy of the various orbitals. The low-energy eigenvalue 
spectrum corresponds to one doubly occupied and one empty $Mo$ orbital 
and singly occupied 
$Cu$ orbitals with all the spin states in this sector being degenerate. In 
each of these spin sectors, we have several states to which approximately 
3eV optical excitation has nonzero transition dipole moment. For a 
reasonable set of parameters the transition dipoles to the S=3 excited 
states are the largest. We also find that the S=3 density of states (DoS) 
is an order of magnitude smaller than the DoS in other spin sectors. This
would lead to long life-time for the S=3 optical excitations resulting in
the system behaving like a S=3 molecular complex after irradiation.

\section{II. Microscopic Model}

\begin{figure}
\includegraphics[height=2.0in,width=2.0in]{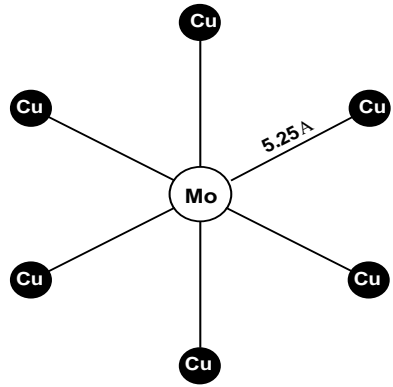}
\caption{\label{figtop} Topology of the $MoCu_6$ system with $Mo$ ion 
at the center and $Cu$ ions at the vertices of a regular hexagon. All the 
$Mo-Cu$ distances are $5.25$ \AA from $Mo$.}
\end{figure}

The topology of the arrangement of the transition metal ions in the 
complex is hexagonal with the $Mo$ ion at the center of the hexagon and 
the $Cu$ ions at the vertices (Fig. \ref{figtop}). The molecular geometry 
is such that there 
is no electron transfer pathway which directly connects the 
copper ions, hence  
all exchange interactions between the $Cu$ ions occur through the 
central $Mo$ ion. Low-symmetry of the complex around the $Mo$ ion implies 
that the degeneracy of the $4d$ orbitals on the $Mo$ ion is lifted. This 
allows one to model photomagnetism in this system by considering only 
two low-lying $4d$ orbitals on $Mo$ and the partly occupied $3d$ orbital on 
each $Cu$. If we restrict the number of $Mo$ orbitals to just one, no
significant exchange interaction between the paramagnetic sites is 
possible since intra-atomic direct exchange interaction would be 
absent. Each $Cu(II)$ ion contributes one electron to the system while the 
$Mo(IV)$ ion contributes two electrons. Thus, we would be dealing with 
an open shell system of eight electrons in eight orbitals. 

\subsection{A. Model parameters}

The essential 
parameters of the model are the effective transfer parameter (the hopping 
integral) between the $Cu$ orbital and each of the two orbitals on $Mo$. It 
is assumed that the hopping integral, $t$, between a $Cu$ orbital and any 
of the two active $Mo$ orbitals is the same. The parameter $\Delta$ 
defines the orbital energy of the lower energy orbital 
on the $Mo$ site, with reference to the orbital energy of the $Cu$ $d$-orbital 
taken to be zero. The energy splitting between the two orbitals on the 
$Mo$ site is given by the 
parameter $\delta$. The quantities $t$, $\Delta$ and $\delta$ are all
parameters of the non-interacting part of the Hamiltonian of the system.

The two-electron integrals, which are important for magnetism and
should appear in the interacting part of the model Hamiltonian are the 
intra-orbital, the on-site inter-orbital
and inter-site inter-orbital electron repulsion integrals. 
The first of these is given by,
\begin{equation}
U^i = [ii|ii] = \int \int \phi_i^*(1) \phi_i(1) \frac{1}{r_{12}}
\phi_i^*(2) \phi_i(2) d^3r_1 d^3r_2
\end{equation}
where $U^i$ is the familiar Hubbard parameter.
We have assumed that the two $Mo$ orbitals have the same Hubbard 
parameter. In cases where a given site has more than one orbital, we need 
to include inter-orbital Coulomb and exchange electron repulsion integrals 
in the model. There are three kinds of inter-orbital repulsion integrals, 
with two orbitals. These are denoted as $U^{ii'}$, $X^{ii'}$ and $W^{ii'}$ 
and are given by,
\begin{eqnarray}
U^{ii'} = [ii|i'i'] = \int \int \phi_i^*(1) \phi_i(1) \frac{1}{r_{12}}
\phi_{i'}^*(2) \phi_{i'}(2) d^3r_1 d^3r_2 \nonumber \\
X^{ii'} = [ii'|ii'] = \int \int \phi_i^*(1) \phi_{i'}(1) \frac{1}{r_{12}}
\phi_i^*(2) \phi_{i'}(2) d^3r_1 d^3r_2 \nonumber \\
W^{ii'} = [ii|ii'] = \int \int \phi_i^*(1) \phi_i(1) \frac{1}{r_{12}}
\phi_i^*(2) \phi_{i'}(2) d^3r_1 d^3r_2 
\end{eqnarray}
where we have used 
the charge cloud notation of chemists' to define the two electron integrals 
both here and elsewhere in the paper \cite{szabo}. 
The single occupancy of the orbitals $\phi_i$ or $\phi_{i'}$, is favoured over 
double occupancy of either, when the inter-orbital repulsion 
$U^{ii'}$ is smaller than the Hubbard parameter $U^i$ or $U^{i'}$ 
for either of the orbitals $\phi_i$ or $\phi_{i'}$. 
The X integral is the exchange integral which favours parallel 
alignment of spins in the orbitals $\phi_i$ and $\phi_{i'}$ when the orbitals 
are singly occupied, leading to the familiar Hund's rule. 
The $W^{ii'}$ term does not play a 
significant role in magnetism but is included in the model for the sake of
consistency since it is larger than the exchange integral. In our model, 
these terms arise only in the case of the $Mo$ atom as it contributes two 
active orbitals to the model, say 1 and 2, thus we have $U^{12}=[11|22]$, 
$W=[11|12]$ and $X=[12|12]$ \cite{ppp}. We also include the extended range electron
correlation corresponding to repulsion between charge densities located 
on different sites, within the zero differential overlap approximation,
via Ohno interpolation \cite{ohno}. The model Hamiltonian can now be 
written as
\begin{eqnarray}
\hat{H} & = & t \sum_{i=3}^8 (\hat E_{1,i} + \hat E_{2,i} + H.c.)   
+\Delta \hat n_1 + (\Delta-\delta)\hat n_2 \nonumber\\
&& + \sum_{i=1}^8 U^i \hat{n}_i(\hat{n}_i-1)/2 + 
U^{12}\hat{n}_1\hat{n}_2 \nonumber \\
&&+ \sum_{i=3}^8 (V_{1i}\hat{n}_1 
\hat{n}_i + V_{2i}\hat{n}_2\hat{n}_i) \nonumber \\
&& + \frac{W}{2}\left[(\hat n_1 + \hat n_2)(\hat E_{12}+\hat E_{21})
 + (\hat E_{12}+\hat E_{21})(\hat n_1 +\hat n_2) \right. \nonumber \\ 
&& \left. -2 (\hat E_{12}+\hat E_{21}) \right] \nonumber \\
&& + \frac{X}{2} (\hat{E}_{12} \hat{E}_{12} + 
\hat{E}_{12} \hat{E}_{21} + H.c + \hat{E}_{21} \hat{E}_{21} \nonumber \\
&& - \hat{n}_{1} - \hat{n}_{2}) 
\end{eqnarray}
Here, H.c. stands for Hermitian conjugate; our numbering scheme
for the orbitals is such that orbitals 1 and 2 are on $Mo$ site and orbitals 
3 through 8 are each located on the copper sites and 
\begin{eqnarray}
\hat{E}_{i,j} = \sum_\sigma \hat{a}_{i,\sigma}^\dagger 
\hat{a}_{j,\sigma};~ \hat{n}_i = \sum_\sigma \hat{a}_{i,\sigma}^\dagger 
\hat{a}_{i,\sigma}
\end{eqnarray}
The inter-site interaction $V_{ij}$ is parametrized as
\begin{eqnarray}
V_{ij} = 14.397\left[ \left(\frac{28.794}{U^i+U^j}\right)^2 + 
r_{ij}^2 \right]^{-\frac{1}{2}}
\end{eqnarray}
with the distance $r_{ij}$ is in \AA ~and energies are in eV. 
The geometry of the cluster $MoCu_6$ is taken to be a regular hexagon 
with a $Mo-Cu$ distance of $5.25$\AA. The topology of the complex used in 
modeling the system is shown schematically in Fig. \ref{figtop}.

\subsection{B. Computational details}

\begin{figure}
\includegraphics[height=3.9in,width=3.2in]{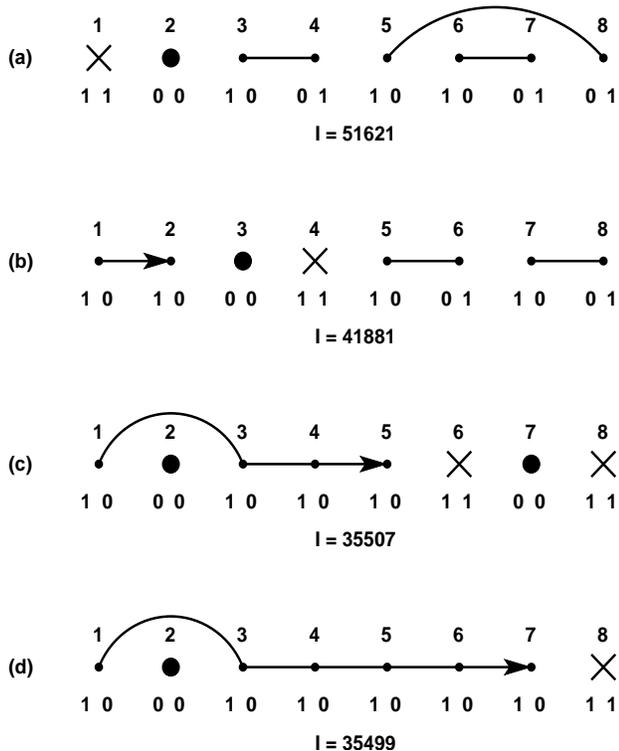}
\caption[]{\label{figvb} Representative VB diagrams in (a) S=0, (b) S=1, 
(c) S=2 and
(d) S=3 subspaces. Dots ($\bullet$)
represent empty sites and crosses ($\times$)
represent doubly occupied sites. In each VB diagram, numbers 1 through 8 
are the site indices and the values at the bottom against each site gives 
the bit state of the bits representing the site in the integer I. 
The bit-pattern and the value of the integer I, representing the VB diagram 
are also shown.}
\end{figure}

The model Hamiltonian spans a finite dimensional Hilbert space 
corresponding to eight electrons in eight orbitals. Since the Hamiltonian 
conserves total spin, the Hilbert space can be further subdivided into 
spaces of total spin 0, 1, 2, 3 and 4 whose dimensions are 1764, 2352, 
720, 63 and 1 respectively. The basis states with a given total spin can 
be generated using a valence bond approach of explicit spin pairings \cite{SR-VB}.
For example, if singly occupied sites $i$ and $j$ are spin paired,
then a line is drawn between sites $i$ and $j$ in the VB diagram 
to indicate the spin pairing, 
$(\alpha_i\beta_j - \beta_i\alpha_j)/\sqrt 2, ~i<j$. We say that the line 
begins at site $i$ and ends at site $j$. If the spins at sites $i_1, i_2 \cdots
i_l$ are not paired, we pass an arrow through these sites in the VB
diagram and this denotes the state $\alpha_{i_1} \alpha_{i_2} \cdots 
\alpha_{i_l}$. Because the Hamiltonian conserves $S^z_{total}$ besides
$S^2$, it is sufficient to work in the subspace $M_S = S$.  
The complete and linearly independent set of VB states in each spin space
can be obtained by taking recourse to modified Rumer-Pauling rules 
\cite{rumer}. Some typical VB diagrams are shown in Fig. \ref{figvb}. 
Each orbital in the VB 
picture has one of four possibilities; the orbital is (i) empty, (ii) 
doubly occupied, (iii) a line begins at the orbital or (iv) a line
ends at the orbital. It is possible 
to associate these four possibilities of an orbital in a VB diagram with 
the four 
states of two bits and stored as sixteen bit integers in an ascending order. 
In any given spin space, the effect of the term $\hat E_{ij}$ in the 
Hamiltonian on a basis state is to alter the orbital occupancy of orbitals 
$i$ and $j$ (subject to Pauli principle) and pair the spins in the orbitals
that were involved to yield a new VB digram with a fixed amplitude. If the 
new VB diagram violates Rumer-Pauling rules, it is trivially possible to 
express it as a linear combination of the basis VB states.

Using the above procedure, the Hamiltonian matrix can be set-up in the 
chosen total spin sector.  The resulting Hamiltonian matrix is sparse 
(since the number of terms in the Hamiltonian is far smaller than the
dimension of the VB complete space) and nonsymmetric matrix and can 
be partially diagonalized to obtain a few of the low-lying eigenstates, 
using Rettrup’s modification of the Davidson algorithm \cite{davidson,rettrup}. 
In some cases, full diagonalization of the matrices is resorted to obtain 
the complete eigenvalue spectrum. In our case, since the 
dimensionalities of the different subspaces are fairly small, we have 
resorted to complete diagonalization of the Hamiltonian matrices in each
spin sector.

In the present problem, the quantities of interest are the optical gaps
in each spin sector and the transition dipoles for these transitions. To
understand the nature of the excited states, it is also important to know 
the spin and charge density distributions in these states. 
Computing many of these 
quantities is simpler in the constant $M_S$ basis. Hence, we transform 
eigenstates in the valence bond basis to the constant $M_S$ basis and carry
out all these calculations.

\section{III. Results and Discussion}

\begin{figure*}
\includegraphics[height=3.0in,width=3.0in]{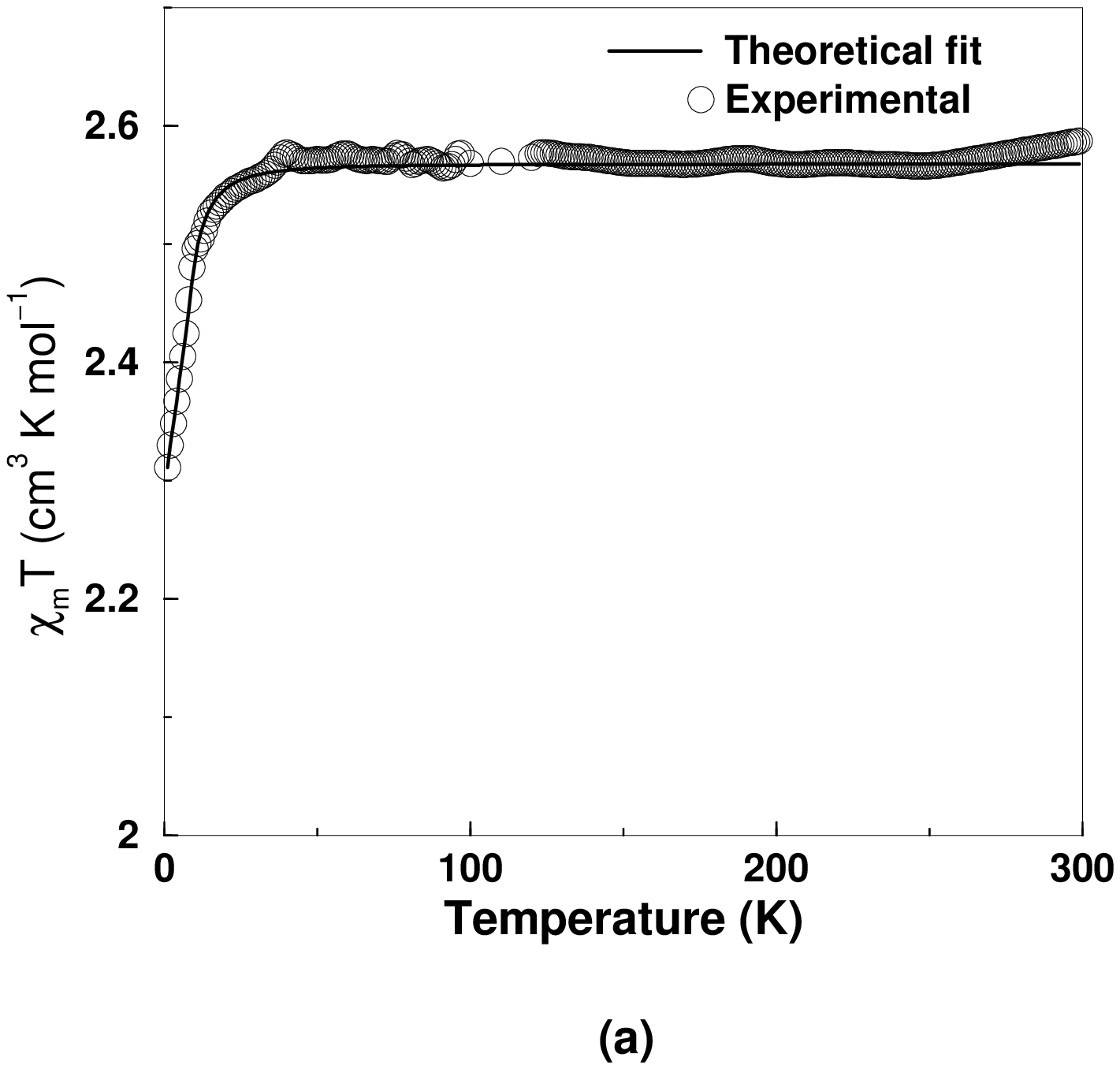}
\hspace*{1cm} \includegraphics[height=3.0in,width=3.0in]{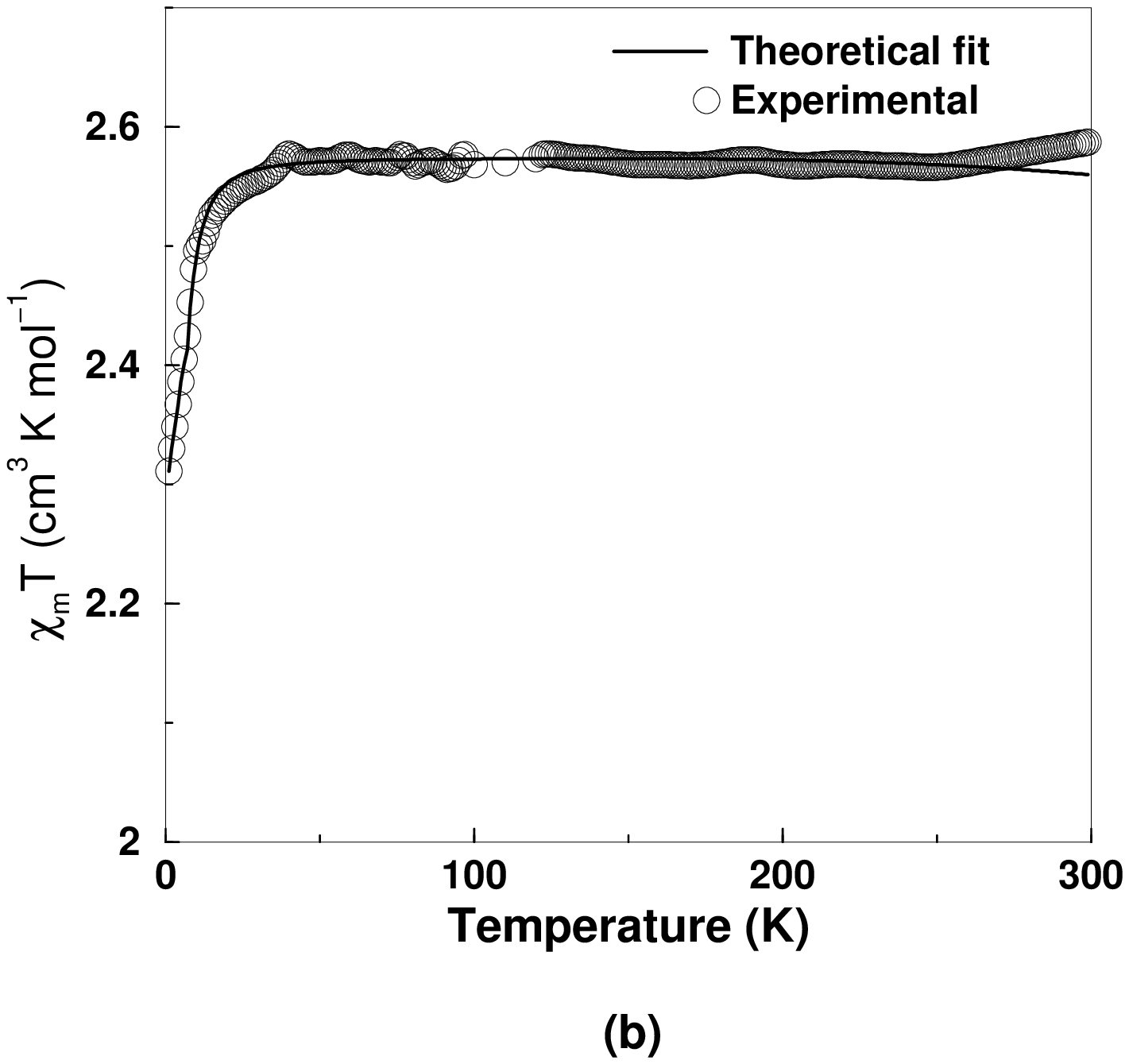}
\caption[]{\label{chit-theo} Theoretical fit of temperature dependence 
of $\chi_MT$ for (a) $t=0.01$ eV and (b) $t=0.00065$ eV. Error in the fit 
$R$ calculated from $R= \sum_i (\chi_{expt}(T_i) - \chi_{calc}(T_i))^2 / 
\sum_i \chi_{expt}(T_i)^2$ for (a) $1 \times 10^{-5}$ and (b) $1.35 
\times 10^{-5}$.}
\end{figure*}
 
We have extensively explored the properties of the model 
over a wide range of parameters, specifically, we have studied the
model at discrete values of the parameters in the following ranges: 
4 eV $ \le U^{Cu} \le$ 6 eV, 0.5 eV $\le U^{Cu}-U^{Mo}\le $ 2.0 eV,
-0.2 eV $\le \Delta \le$ -0.5 eV, 0.0 eV $\le \delta \le$ 0.3 eV, 
0.5 eV $\le U^{Mo} - U^{Mo-Mo} \le $ 1.5 eV, 0.5 eV $ \le W \le $ 1.5 
eV with $X \le W/2$. However, we report our results for the model 
parameters values of $U^{Cu}$=6.0 eV, $U^{Mo}$=4.0 eV, $\Delta$=-0.4 eV, 
$\delta$=0.0 eV, $U^{Mo-Mo}$ = 3.5 eV, $W$=0.75 eV and $X$=0.3 eV since
it gives a reasonable value for the optical gap of the system. 
Our assumption of $U^{Cu}$ being higher than $U^{Mo}$ is justified 
since $3d$ orbitals are more compact than the $4d$ orbitals. The interelectron
repulsion integral involving the d orbitals on $Mo$, ($U^{Mo-Mo}$) is slightly
less than the intra-orbital repulsions. We expect the integral $W$ to be 
larger than $X$, and both are expected to be significantly smaller than 
the $U$ type repulsion integrals that involve repulsion between two 
electron densities, each one located in a single orbital. The orbital 
energy of the $Mo$ $4d$ orbitals could be slightly smaller than the $Cu$ $3d$ 
orbital energy since $Mo$ has a higher atomic number. Therefore, we have
assumed $\Delta$ to be  small and negative. These values of the interaction 
parameter are reasonable as found from theoretical studies on transition 
metal oxides in the context of high-T$_c$ superconductors \cite{DD}. 
For these 
model parameters values, transfer integrals in the range 6x10$^{-4}$ eV 
$ \le t \le $ 0.01 eV fit the experimental $\chi T$ {\it vs} $T$ 
data. This is because, the various spin states with all Copper ions in 
+2 oxidation state and Molybdenum ion in +4 oxidation state are degenerate
to within about $t^2/U$, where U is an average Hubbard parameter ($\sim$ 
5 eV). The near degeneracy of these "covalent" spin states ensures that
the behaviour for $T > t^2/U$ corresponds to that of six free
spins on the $Cu(II)$ ions. We have fixed the value of $t$ at 0.01 eV to
ensure that there is a significant transition dipole between the 
low-lying spin state and its optically coupled state.

\subsection{A. Low-energy spectrum of the model}

To understand the eigenvalue spectrum of the model, it is instructive to 
analyze the $t=0$ behaviour. In this case, the eigenstates of the model can 
be completely classified based on the site occupancies. The only off-diagonal
contribution that mixes the states with different occupancies arises from 
the $W$ and the $X$ terms involving the $Mo$ orbitals. These terms, however, 
conserve the total number of electrons on $Mo$. When we have the $4d^2$ 
configuration on $Mo$ and all the copper ions in $3d^9$ configuration, we 
have a total of 384 states. This corresponds to six ways of occupying two 
electrons in the two $4d$ orbitals and sixty four spin configurations of the 
six spins on the six $Cu(II)$ ions. For the model employed by us, the lowest 
energy state corresponds to the lower of the orbitals on $Mo$ being doubly 
occupied, for $U^{Mo} < (U^{Mo-Mo} + \delta)$, ignoring the off-diagonal
contributions from the $W$ and $X$ terms. The effect of the off-diagonal 
$W$ and $X$ terms is (i) to favour double occupancy on the lower energy 
$Mo$ orbital and (ii) to favour parallel spin alignment when these orbitals 
are singly occupied. Corresponding to this occupancy, five states with 
total spin S=0, nine states with total spin S=1, five states with total 
S=2 and one state with total spin S=3 are all degenerate. There is only 
one S=4 state which is 0.671 eV above the ground state and corresponds 
to single occupancy of each orbital with all spins aligned in parallel. 
These states span all the 64 spin configurations corresponding to six spin-1/2 
objects; since all these states are degenerate to within a few degree K, 
the paramagnetic behaviour observed in the system before irradiation is 
reproduced for small values of $t$. The theoretical fit to experimental 
data for two different values of the transfer integral is shown in 
Fig. \ref{chit-theo}.

The next set of states corresponds to single occupancy of each of the $Mo$ 
orbitals. If the spins of the electrons in these orbitals are parallel, 
the states will have a lower energy than if they are antiparallel. There
are in all 256 such states of which 128 states have parallel spin alignment
and another 128 states have antiparallel spin alignment. The remaining
sixty four of the 384 states have a double occupancy in the higher
energy $Mo$ orbital. These 384 states form the low energy spectrum when
$U^{Cu}$ is significantly larger than $(U^{Mo}-\Delta)$ so that the intersite
interactions do not alter the picture. The most important point to note
is that since the low-energy states all have identical charge distribution,
the transition dipole matrix elements between these states are zero and 
this result holds even when $t \ne 0$.

\subsection{B. Charge-Transfer Excited States}
\begin{figure}[h]
\includegraphics[height=4.5in,width=3.0in]{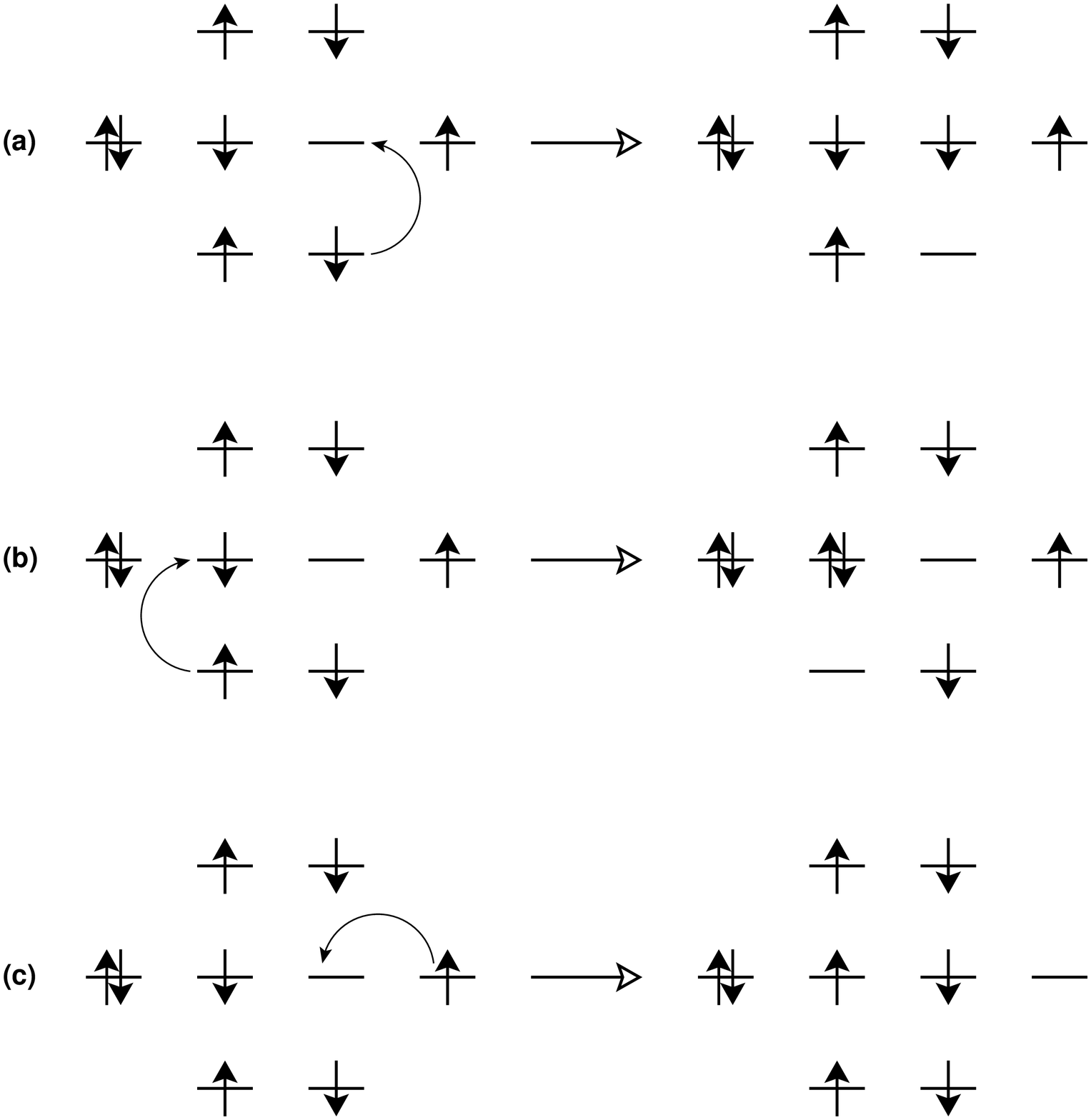}
\caption{\label{figexcited} The virtual excited state resulting from
the electron transfer shown in (a) has a lower energy than those
shown in (b) and (c), because of the Hund's rule. The superexchange 
process in (a) favours S=3 state, and in (b) an S=2 state. The process 
(b) is preferred over (a) if the splitting of the orbitals
on the $Mo$ site as well as $U^{Mo-Mo}$ are large. In all the figures, 
the two orbitals in the center are the $Mo$ orbitals and the six 
peripheral orbitals are the $Cu$ orbitals.}
\end{figure}

The higher energy excited states of the system involve 
charge-transfer (CT) from the $Mo(IV)$ ion to $Cu(II)$ resulting in a state 
with a large weight for the $Mo(V)-Cu(I)$ configurations. 
These states are found 
at about 3 eV above the ground state. These excitations have nonzero 
transition dipole to the ground state. One of the most important features of 
this excitation is that, in the excited state there is considerable weight 
for the basis state with singly occupied $Mo$ site. The singly occupied 
$Mo(V)$ site favours virtual electron transfer from the $Cu(II)$ sites, of 
electrons which have the same spin orientation as the unpaired $Mo(V)$ 
electron, because of the Hund's rule. The virtual transfer of an electron 
from $Cu(II)$ with its spin orientation same as that already present on 
$Mo(V)$ is favoured as long as (i) the $d$-orbitals on $Mo$ 
are nearly degenerate 
and (ii) the inter-orbital $d-d$ repulsion, on $Mo$ is weaker than the 
intra-orbital repulsion ($U^{Mo-Mo} < U^{Mo}$). This feature contributes 
to stabilizing the high-spin CT excitation relative to the 
low-spin excitations. In Fig \ref{figexcited} we show the possible pathways for 
superexchange in this state. The factor which favours stabilization of 
the low-spin state is the large number of pathways that exist for the 
delocalization of electrons in the low-spin state, compared to the 
high-spin states. In our case, there are 1764 singlets and 2352 triplets 
while there are only 63 S=3 states. The splitting of the states with 
different spins in the CT state is thus a competition between 
direct exchange which stabilizes the high-spin state and the kinetic 
exchange which favours the low-spin state due to the availability of a 
larger phase space for electron delocalization. This implies that the 
lowest-energy CT excited state could have any spin, $S$, in the range
0 $\le$ S $\le$ 3. 

\begin{figure}[]
\includegraphics[height=3.5in,width=3.5in]{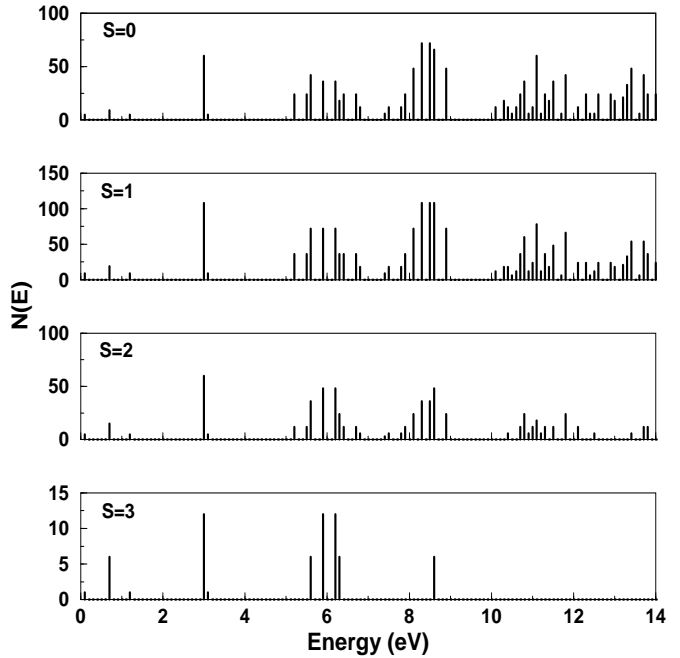}
\caption{\label{figdos} Density of states versus energy in S=0,1,2 and 3 
subspaces. There is only one state in S=4 sector at 0.6709 eV from the 
ground state (not shown in the figure).}
\end{figure}

\begin{table*}
\noindent
\caption{\label{tabtdip} Energy gaps (in eV) to the charge-transfer 
excited states
from the lowest state in different spin sectors and the magnitude of 
the corresponding transition dipole moments in $10^{-3}$ Debye.
Also presented in the table are the total spin and charge densities
on the Molybdenum site in the excited state.}
\begin{center}
\begin{tabular}{|c|c|c|c|c|c|} \hline
Spin sector & Energy gap (eV) & Degeneracy & Transition dipole & $<S_z^{Mo}>$ & 
$<n_{Mo}>$ \\
& & &(x$10^{-3}$ Debye)& &\\\hline
S=0 & 2.9921 & 6 & 7.0 & 0.0000 & 1.0000\\
    & 2.9923 & 4 & 4.1 & 0.0000 & 1.0000\\\hline

    & 2.9920 & 4 & 6.2 &-0.2498 & 1.0000\\
S=1 & 2.9923 & 8 & 5.3 & 0.4986 & 1.0000\\
    & 2.9924 & 2 & 3.4 & 0.5000 & 1.0000\\\hline

    & 2.9921 & 2 & 8.3 &-0.1596 & 1.0000\\
S=2 & 2.9923 & 2 & 12.1& 0.5000 & 1.0000\\
    & 2.9924 & 4 & 7.8 & 0.5000 & 1.0000\\ \hline

S=3 & 2.9925 & 5 & 16.0 & 0.5000 & 1.0000\\ \hline
\end{tabular}
\end{center}
\label{ctxtns}
\end{table*}

If the splitting of the $Mo$ $d$ orbitals is large or $U^{Mo-Mo}$ is large, 
then the intermediate virtual state with a doubly occupied $d$ orbital on 
$Mo$ is favoured, resulting in an antiferromagnetic interaction between 
the $Cu(II)$ ion and the $Mo$ spin (Fig \ref{figexcited}). This factor would contribute 
to a lowering of the S=2 CT excited state relative to 
CT excited states with spin S$\ne$2. However, the final 
outcome depends on both the direct exchange and the kinetic exchange 
contributions. Thus, it is again difficult to predict which of the spin 
states is a low-energy CT excited state. We see all the above 
scenarios in our calculations, when we widely vary the model parameters.

The above picture is slightly altered by the long-range interactions in
the model. These interactions depend upon the occupancies of the $Mo$ site
as well as all the $Cu$ sites and are significant since the molecular complex 
is insulating and screening of these interactions is not significant. 
Together with a nonzero transfer integral, the long-range interactions 
complicate the picture while retaining the general 
physical features outlined above. 

For the model parameter for which we are reporting the results, 
we show the histogram giving the number of states
as a function of energy in Fig. \ref{figdos} for the full spectrum of 
the model in each total spin sector. We find CT excitations at about 
2.993 eV in all the spin sectors. The energies and magnitude of the 
transition dipole vectors from the ground state are given in Table 
\ref{tabtdip}. We also present in Table \ref{tabtdip}, the degeneracy of 
states as well as the charge and spin density on the $Mo$ site. We 
note that in all the states located at this energy, independent of the 
total spin, the charge density on the $Mo$ site is one. What is interesting 
is that while the energies of these states as well as the charge density on 
the $Mo$ site are all the same, the transition dipole moment is largest 
for the S=3 states. This fact, coupled with the high degeneracy of the state 
implies that the S=3 states have the largest absorption to the CT excited 
state.

An important difference between the S=3 eigenvalue spectrum and eigenvalue
spectra of other spin states is that the DoS in the former
is very dispersed and the states are far apart in energy, when not degenerate.
This has implications for the life-time of the excited states. In the S=0,
1 and 2 spaces, there are an order of magnitude more states near the 
CT excitation gap of 2.993 eV compared with the S=3 states
at this energy. Thus, other spin states have pathways for internal conversion
leading to rapid de-excitation of the CT state. This is less
likely in the S=3 state. For this reason, the S=3 excited state has a 
longer life-time and can be observed in the photomagnetism experiments. One
feature that is however missed in our studies is the computation of the
equilibrium geometry in the CT excited state specifically around the $Cu(I)$ 
site that may play a crucial role in increasing the life-time of 
the state. This is rather
hard to incorporate in our model wherein the ligand atoms are completely
neglected. It is possible that the CT excited state with 
S=3 is more distorted (relative to the ground state) than similar states
with different total spin. This could further enhance the life-time of the
S=3 CT excitation.

For the sake of completeness, it is worth mentioning that there are also
different kinds of charge transfer excited states at much higher energies
corresponding to electron transfer from the $Cu(II)$ site to the $Mo(IV)$ site,
charge transfer between $Cu(II)$ sites and so forth for which the transition
dipole moments are nonzero. However, these states occur at a much higher 
energy and are not excited by laser light at 3eV.

The model developed in this paper is specific to the isolated molecular
complex we have studied. However, this model also throws light on the
mechanism leading to the observed  
ferromagnetism in the PIPT systems. While the degeneracy of the
different optically excited spin states is not significantly affected
by the exchange mechanism operative in the optically excited state, the
long life-time of the high-spin state could be responsible for the
observed ferromagnetism in extended systems.

\section{IV. Conclusions}

In this paper, we have introduced a model for the photomagnetism 
in the heptanuclear complex, $[Mo(IV)(CN)_2(CN-CuL)_6]^{8+}$ perchlorate
which involves the six partly filled $3d$ orbital on the $Cu(II)$ sites and
two lowest energy $4d$ orbitals on the $Mo(IV)$ site. Besides the site energies
of the orbitals and the transfer integrals between the $Cu$ and $Mo$ orbitals,
we introduce on-site and extended range interactions between the various
sites. On the $Mo$ site, we consider all the three types of two 
electron integrals involving the $Mo$ orbitals. Using this model, we compute
the full eigenvalue spectrum, the transition dipoles with the ground state, spin
and charge densities for all the states in each of the spin 
sectors by exactly solving the model in a spin conserving valence bond basis.
Our studies show that the ground state is sixty four fold degenerate with an
unpaired electron on each $Cu$ site. This leads to the observed paramagnetic
susceptibility prior to irradiation, of six non-interacting spin-1/2 objects
per formula unit.
The excited states to which there is nonzero transition dipole in each spin
sector are located at about 2.993 eV (for the values of the parameter we have
chosen) in each of the spin sectors. The number of states at this energy as
well as the transition dipole moment for the transition are highest for the
S=3 states. There also exist very few states in the S=3 sector below this 
energy, thereby introducing a bottle-neck for its internal conversion for 
non-radiative decay to the ground state. There may also occur changes 
in the geometry of the complex in the excited state which renders a low
Frank-Condon factor for radiative decay. Thus, the observed post radiation
magnetic behaviour of the complex with 75\% of the molecules in the S=3 state 
can be rationalized by our model.

\section{Acknowledgment}

The authors thank DST for the support received under a joint Indo-French
Laboratory for Solid State Chemistry (IFLaSC) and Indo-French Centre for
Promotion of Advanced Research (IFCPAR) for generous support under 
Project 3108-3 on Design, Synthesis and  Modeling Molecular Magnets.
The authors also thank the French ministry of research (ACI young researcher 
No JC4123) and the European Community (Network of Excellence, MAGMANet) for 
the financial support.

\end{document}